\documentclass[runningheads,fleqn]{svmult}

\usepackage{subfigure}
\usepackage{epsfig}
\usepackage{psfrag}
\usepackage{rotating}
\usepackage{axodraw}

\usepackage{makeidx}   
\usepackage{graphicx}  
\usepackage{subeqnar}  
\usepackage{multicol}  
\usepackage{physmult}  
\makeindex             

\newcommand{\vcb}{|V_{cb}|}

\newcommand{\vub}{|V_{ub}/V_{cb}|}
\newcommand{\vts}{|V_{ts}|}

\def\epe{\varepsilon'/\varepsilon}

\newcommand{\gev}{\, {\rm GeV}}

\newcommand{\bea}{\begin{eqnarray}}
\newcommand{\eea}{\end{eqnarray}}
\newcommand{\bd}{\begin{displaymath}}
\newcommand{\ed}{\end{displaymath}}
\newcommand{\beq}{\begin{equation}}
\newcommand{\eeq}{\end{equation}}
\newcommand{\be}{\begin{equation}}
\newcommand{\ee}{\end{equation}}
\newcommand{\bi}{\begin{itemize}}
\newcommand{\ei}{\end{itemize}}
\newcommand{\ord}{{\cal O}}

\def\kpn{K^+\rightarrow\pi^+\nu\bar\nu}

\def\klpn{K_{\rm L}\rightarrow\pi^0\nu\bar\nu}
\newcommand{\kmm}{K_{\rm L} \to \mu^+ \mu^-}

\begin{document}
%


\thispagestyle{empty}
\phantom{xxx}
\vskip1truecm
\begin{flushright}
 TUM-HEP-544/04 \\
 hep-ph/0402191 \\  
February 2004
\end{flushright}

\vskip1.4truecm

\begin{center}
\boldmath
{\LARGE{\bf  Waiting for Clear Signals  of New Physics
             in B and K Decays}}

\unboldmath
\end{center}

   \vskip0.6truecm
\centerline{\Large\bf Andrzej J. Buras}
\bigskip
\centerline{\sl Technische Universit{\"a}t M{\"u}nchen}
\centerline{\sl Physik Department} 
\centerline{\sl D-85748 Garching, Germany}
\vskip1truecm
\centerline{\bf Abstract}

We classify the extensions of the Standard Model (SM) according to the 
structure of local operators in the weak effective Hamiltonian and the
presence or  absence of new flavour and CP-violating interactions 
beyond those represented by the CKM matrix. In particular we review
characteristic properties of models with minimal flavour violation (MFV), 
models
with significant contributions from Higgs penguins and models with enhanced
$Z^0$ penguins carrying a large new CP-violating phase. Within the latter
models, the anomalous behaviour of certain $B\to\pi K$ observables implies
large departures from the SM predictions for rare and CP-violating $K$ and 
$B$ decays. Most spectacular is the enhancement of $Br(\klpn)$ by one order 
of magnitude and a strong violation of the MFV relation
$(\sin2\beta)_{\pi\nu\bar\nu}=(\sin2\beta)_{\psi K_S}$.
 On the other hand 
our prediction for $(\sin2\beta)_{\phi K_S}\approx 0.9$ differs from the
Belle result by the sign but is consistent with the BaBar value. 
We give a personal shopping list for the coming years.

\vskip1.2truecm

\centerline{\it  Invited Talk given at }
\centerline{\bf the 9th Adriatic Meeting}
\centerline{\bf ``Particle Physics and the Universe"}
\centerline{\it Dubrovnik, September 4 -- 14, 2003}

\newpage
\setcounter{page}{0}

\title*{Waiting for Clear Signals of New Physics in B and K Decays}
\toctitle{Waiting for Clear Signals of New Physics in B and K Decays}
%
%
\titlerunning{Waiting for Clear Signals of New Physics\\ in B and K Decays}
%
\author{Andrzej J. Buras}
\authorrunning{Andrzej J. Buras}
%
%

\maketitle              
\noindent
Technical University Munich, Physics Department\\
D-85748 Garching, Germany


\section{Introduction}
The quark flavour dynamics of the Standard Model (SM) is consistent
with the existing data on $K$ and $B$ meson decays within experimental and
theoretical uncertainties. In spite of this, most of
us expect that when the precision of experiments and also the 
theoretical tools
improve, some clear signals of new physics (NP) at very short distance scales
will
be seen. As the search for new phenomena through $K$ and $B$ decays is
necessarily an indirect one, it will not be easy to find out what precisely
this NP is. This will be even the case in the presence of clear
deviations from the SM expectations. 

Yet, remembering that decays like
$K_L\to\mu^+\mu^-$, $K_L\to\pi\pi$ and the $K^0-\bar K^0$ mixing played an
important role in the construction of the SM \cite{Weinberg:2004kv} 
and led to the GIM mechanism
\cite{GIM}  and the CKM matrix \cite{CKM}, we are confident that these 
indirect signals of NP
will give us definitive hints where to go and where not to go. This in turn
will allow us to select few competing theories which will be tested
directly through high energy collider experiments. Moreover, these signals
will hopefully give us additional hints for the fundamental flavour dynamics 
at ultra 
short distance scales that cannot be tested directly in a foreseeable future.
It is  obvious that for this project to become successful, it is
essential to
\begin{itemize}
\item
make the SM predictions for $K$ and $B$ decay observables as accurate as
possible, in order to be sure that the observed deviations from the SM
expectations originate in NP contributions and are not due to
our insufficient understanding of hadron dynamics and/or truncation of the
perturbative series.
\item
consider
simultaneously as many processes as possible. Only in this manner the
paramaters of a given theory can be fully determined and having them at hand
predictions for other observables can be made. In this enterprise
correlations between various observables play an important r\^ole, as they 
may exclude or pinpoint a given extension of the SM even without a detailed
knowledge of the parameters specific to this theory.
\item
consider observables which while being sensitive to the short distance
structure of the theory are only marginally sensitive to the long distance
dynamics that is not yet fully under control at present.
\end{itemize}

These notes review  selected aspects of this program. Due to severe space 
limitations I will follow the strategy of many of my colleagues and will
concentrate to a large extent on my own work and the work
done in my group in Munich. In Section 2, after
giving a master formula for weak decay amplitudes, I will group various
extentions of the SM in five classes. In subsequent three sections, I will
discuss the first three classes that, being more predictive than the remaining
two, should be easiest to test. 
A shopping list in Section 6, with the hope to be able to distinguish between
various NP scenarios, ends this presentation.
Recent more detailed reviews can be found in \cite{REV}.

\section{Theoretical Framework}
        \label{sec:TH}
\subsection{Master Formula for Weak Decays} 
The present framework for weak decays is based on the operator product 
expansion that allows to separate short ($\mu_{SD}$) and long 
 ($\mu_{LD}$) distance 
contributions to weak amplitudes and on the renormalization group (RG) 
methods that allow to sum large logarithms $\log \mu_{SD}/\mu_{LD}$  to 
all orders in perturbation theory. The full exposition of these methods 
can be found in \cite{AJBLH,BBL}.

An amplitude for a decay of a meson 
$M= K, B,..$ into a final state $F=\pi\nu\bar\nu,~\pi\pi,~\pi K$ is then
simply given by
\be\label{amp5}
A(M\to F)=
\frac{G_F}{\sqrt{2}}\sum_i V^i_{CKM}C_i(\mu)\langle F|Q_i(\mu)|M\rangle.
\ee
Here $G_F$ is the Fermi constant and $Q_i$ are the relevant local
operators which govern the decays in question. 
They are built out of quark and lepton fields.
The Cabibbo-Kobayashi-Maskawa
factors $V^i_{\rm CKM}$ \cite{CKM}, 
the matrix elements $\langle F|Q_i(\mu)|M\rangle$
and the Wilson coefficients $C_i(\mu)$, evaluated at the
renormalization scale $\mu$, describe the 
strength with which a given operator enters the amplitude.

Formula (\ref{amp5}) can be cast into 
 a master formula for weak decay amplitudes that 
 goes beyond the SM \cite{Pisa}. 
It reads (we suppress $G_{\rm F}$):
\begin{eqnarray}\label{master}
{\rm A(Decay)} &=& \sum_i B_i \eta^i_{\rm QCD}V^i_{\rm CKM} 
\lbrack F^i_{\rm SM}+F^i_{\rm New}\rbrack
\nonumber\\
& &  +
\sum_k B_k^{\rm New} \lbrack\eta^k_{\rm QCD}\rbrack^{\rm New} V^k_{\rm New} 
\lbrack G^k_{\rm New}\rbrack\, .
\end{eqnarray}
The non-perturbative parameters $B_i$ represent the matrix elements of local 
operators present in the SM. For instance in the case of 
$K^0-\bar K^0$ mixing, the matrix element of the operator
$\bar s \gamma_\mu(1-\gamma_5) d \otimes \bar s \gamma^\mu(1-\gamma_5) d $
is represented by the parameter $\hat B_K$.
There are other non-perturbative parameters in the SM that represent 
matrix elements of operators $Q_i$ with different colour and Dirac 
structures. The objects $\eta^i_{\rm QCD}$ are the QCD factors resulting 
from RG-analysis of the corresponding operators and $F^i_{\rm SM}$ stand for 
the so-called Inami-Lim functions \cite{IL} that result from the calculations 
of various
box and penguin diagrams in the SM. They depend on the top-quark mass. 
 
New physics can contribute to our master formula in two ways. It can 
modify the importance of a given operator, present already in the SM, 
through the new short distance functions $F^i_{\rm New}$ that depend on 
the new 
parameters in the extensions of the SM like the masses of charginos, 
squarks, charged Higgs particles and $\tan\beta=v_2/v_1$ in the MSSM. 
These new 
particles enter the new box and penguin diagrams. In more complicated 
extensions of the SM new operators (Dirac structures), that are either 
absent or very strongly suppressed in the SM, can become important. 
Their contributions are described by the second sum in 
(\ref{master}) with 
$B_k^{\rm New}, \lbrack\eta^k_{\rm QCD}\rbrack^{\rm New}, V^k_{\rm New}, 
G^k_{\rm New}$
being analogs of the corresponding objects in the first sum of the master 
formula. The $V^k_{\rm New}$ show explicitly that the second sum describes 
generally new sources of flavour and CP violation beyond the CKM matrix. 
This sum may, however, also include contributions governed by the CKM 
matrix that are strongly suppressed in the SM but become important in 
some extensions of the SM. In this case $V^k_{\rm New}=V^k_{\rm CKM}$. 
Clearly the new functions $F^i_{\rm New}$ and $G^k_{\rm New}$ as well as the 
factors $V^k_{\rm New}$ may depend on new CP violating phases complicating 
considerably phenomenological analyses.

\subsection{Classification of New Physics}
Classification of new physics contributions can be done in various ways. 
Having the formula (\ref{master}) at hand let us classify
these contributions from the point
of view of the operator structure of the effective weak Hamiltonian,
the complex phases present in the Wilson coefficients of the
relevant operators and the distinction whether the flavour changing
transitions are governed by the CKM matrix or by new sources of
flavour violation \cite{Erice}.
For the first four classes below we assume that there are only three
generations of quarks and leptons. The last class allows for
more generations.

{\bf Class A}

This is the simplest class to which also the SM belongs.
In this class there are no new complex phases and flavour changing 
transitions are  governed  by the CKM matrix.
Moreover, the only relevant operators are those that are relevant in the SM.
Consequently NP enters only through the Wilson coefficients of the SM 
operators that can receive new contributions
  through diagrams involving new internal particles.

The models with these properties will be called  
Minimal Flavour Violation (MFV) models, as defined in 
\cite{UUT}. Other definitions can be found in \cite{AMGIISST,BOEWKRUR}. 
In this case  our master formula simplifies to 
\be\label{mmaster}
{\rm A(Decay)}= \sum_i B_i \eta^i_{\rm QCD}V^i_{\rm CKM} 
 F_i(v)
\qquad
 F_i=F^i_{\rm SM}+F^i_{\rm New}~~{\rm (real),}
\ee 
where $F_i(v)$ are the {\it  master functions} of MFV models \cite{Cracow}
\be\label{masterf}
S(v),~X(v),~Y(v),~Z(v),~E(v),~ D'(v),~ E'(v)
\ee
with $v$ denoting collectively the parameters of a
given MFV model. A very detailed account of  MFV can be found in
\cite{Cracow}. In Section 3 some of its main features will be recalled.
Examples of models in this class are the Two Higgs Doublet Model II and 
the constrained MSSM
if $\tan\beta$ is not too large. Also models with one extra universal 
dimension are of MFV type.

{\bf Class B}

This class of models differs from class A through the contributions
of new operators not present in the SM. It is assumed, however,
that no new complex phases
beyond the CKM phase are present. We have then
\begin{eqnarray}\label{masterB}
{\rm A(Decay)} &=& \sum_i B_i \eta^i_{\rm QCD}V^i_{\rm CKM} 
\lbrack F^i_{\rm SM}+F^i_{\rm New}\rbrack
\nonumber\\
& &  +
\sum_k B_k^{\rm New} \lbrack\eta^k_{\rm QCD}\rbrack^{\rm New} V^k_{\rm CKM} 
\lbrack G^k_{\rm New}\rbrack\,
\end{eqnarray}
with all the functions $F^i_{\rm SM}$, $F^i_{\rm New}$ and $G^k_{\rm New}$ 
being {\rm real}. 
Typical examples of new Dirac structures are the 
operators $(V-A)\otimes(V+A)$, 
$(S-P)\otimes (S\pm P)$ and 
$\sigma_{\mu\nu} (S-P) \otimes \sigma^{\mu\nu} (S-P)$ contributing to 
$B_{d,s}^0-\bar B^0_{d,s}$ mixings that become relevant in the MSSM 
with a large $\tan\beta $. 

{\bf Class C}

This class of models differs from class A through the presence of
new complex phases in the Wilson coefficients of the usual SM
operators. Contributions of new operators can
be, however, neglected.  An example is the
MSSM with  not too a large $\tan\beta$ and with non-diagonal elements
in the squark mass matrices.
This class can be summarized by
\be\label{mmaster3}
{\rm A(Decay)}= \sum_i B_i \eta^i_{\rm QCD}V^i_{\rm CKM} 
 F_i(v)
\qquad
 F_i(v)~~{\rm (complex).}
\ee

 {\bf Class D}

 Here we group models with new complex phases, new operators
 and new flavour changing contributions which are not governed
 by the CKM matrix. As now the amplitudes are given by the most general
 expression (\ref{master}), the phenomenology in this class of models
 is more involved than in the classes A--C \cite{MPR,GGMS}.
  Examples of models in class D are multi-Higgs models 
 with complex phases in the  Higgs sector, general SUSY models, 
 models with spontaneous
 CP violation and left-right symmetric  models.

 {\bf Class E}

 Here we group  models in which
the unitarity of the three generation CKM matrix does not
hold. 
Examples are four generation models and models with tree
level FCNC transitions. If this type of physics is present,
the unitarity triangle does not close. The most recent discussion of the 
possible violation of the unitarity of the three generation CKM matrix can be
found in \cite{UNIT}.

\section{Class A: MFV Models}

\subsection{Model Independent Relations}
One can derive a number of relations between various observables that 
do not depend on the functions $F_i(v)$ and consequently are universal 
within the class of the MFV models. 
Violation of any of these relations would be a signal of the presence of 
new operators and/or new weak complex phases.
We list here the most interesting relations of this type.

\begin{figure}[htb!]
\begin{center}
{\epsfig{figure=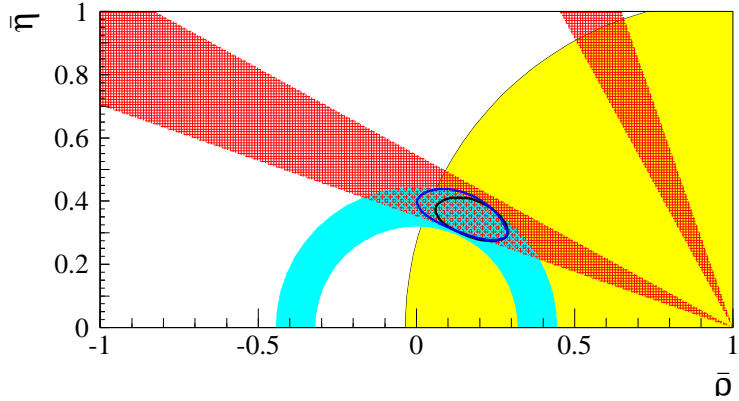,height=6cm}}
\caption[]{The allowed 95$\%$ regions in the 
$(\bar\varrho,\bar\eta)$ plane in the SM (narrower region) and in the 
MFV models (broader region) from the update of \cite{BUPAST}.
The individual 95$\%$ regions for the constraint from 
$R_b$, $\Delta M_d/\Delta M_s$ and $\sin 2 \beta$ are also shown.  
}
\label{fig:figmfv}
\end{center}
\end{figure}

{\bf 1.} A universal unitarity triangle (UUT) common to all MFV models can be
constructed by using only observables that are independent of $F_i(v)$
\cite{UUT}. 
In particular its two sides $R_b$ and $R_t$ \cite{REV} can be determined 
from $\vub$ and
the ratio of $B^0_{d,s}-\bar B^0_{d,s}$ mixing mass differences $\Delta
M_{d,s}$, respectively:
\be\label{RTVUB}
R_b=4.4\left|\frac{V_{ub}}{V_{cb}}\right|,\quad
R_t=0.90~\left[\frac{\xi}{1.24}\right] \sqrt{\frac{18.4/ps}{\Delta M_s}} 
\sqrt{\frac{\Delta M_d}{0.50/ps}},
\ee
where $\xi=1.24\pm 0.08$ \cite{CERNCKM} is a non-perturbative parameter,
$\Delta M_d=(0.503\pm0.006)/{\rm ps}$ and $\Delta M_s>14.4/{\rm ps}$.
Moreover, the angle {$\beta$} of this triangle can be found from
measurements of the time-dependent CP asymmetry
$a_{\psi K_S}(t)$ with the result \cite{BaBar,Belle}
\be
(\sin 2\beta)_{\psi K_S}=0.736\pm 0.049~.
\label{ga}
\ee
Using (\ref{RTVUB}) and (\ref{ga}) one finds the apex of the UUT 
placed within the larger ellipse in  fig.~\ref{fig:figmfv} \cite{BUPAST}.  
A similar analysis has been performed in \cite{AMGIISST}.

{\bf 2.} Two theoretically  clean relations are \cite{AJB03} 
\begin{equation}\label{R1}
\frac{Br(B\to X_d\nu\bar\nu)}{Br(B\to X_s\nu\bar\nu)}=
\left|\frac{V_{td}}{V_{ts}}\right|^2 ,
\qquad
\frac{Br(B_{s}\to\mu\bar\mu)}{Br(B_{d}\to\mu\bar\mu)}
=\frac{\hat B_{d}}{\hat B_{s}}
\frac{\tau( B_{s})}{\tau( B_{d})} 
\frac{\Delta M_{s}}{\Delta M_{d}}.
\ee
They do not 
involve the $B_q$--meson decay constants $F_{B_q}$ and consequently contain 
substantially smaller hadronic uncertainties than the formulae 
for individual branching ratios \cite{Cracow}.
The ratio $\hat B_{s}/\hat B_{d}$
is known from lattice calculations 
with a respectable precision \cite{CERNCKM}:
\be\label{BBB}
\frac{\hat B_{s}}{\hat B_{d}}=1.00\pm 0.03, \qquad
\hat B_{d}=1.34\pm0.12, \qquad \hat B_{s}=1.34\pm0.12~.
\ee
With a future precise measurement of $\Delta M_s$, the second 
formula in (\ref{R1}) 
will give a  precise prediction for the ratio 
of the branching ratios $Br(B_{q}\to\mu\bar\mu)$.

{\bf 3.} It is possible to derive an
accurate formula for $\sin 2\beta$ that depends only on the 
$K\to\pi\nu\bar\nu$ branching 
ratios and a calculable $\bar P_c(X)=0.38\pm0.06$ 
\cite{BBSIN,BFRS-III,BFRS-II}:
\be\label{sin2bnunu}
\sin2(\beta-\beta_s)= \frac{2 r_s}{1+r_s^2}, \qquad 
r_s=\frac{\varepsilon_1\sqrt{B_1-B_2}-\bar P_c(X)}
{\varepsilon_2\sqrt{B_2}}\,,
\ee 
where $\beta_s\approx -1^\circ$ enters $V_{ts}=-|V_{ts}|\exp(-i\beta_s)$, 
$\varepsilon_i=\pm 1$ and
\begin{equation}\label{b1b2}
B_1={Br(\kpn)\over 4.78\cdot 10^{-11}},\qquad
B_2={Br(\klpn)\over 2.09\cdot 10^{-10}}.
\end{equation}
In the MFV models $\varepsilon_1=\varepsilon_2={\rm sgn}(X)$ \cite{BF01}, 
where $X$ is the relevant master function.

{\bf 4.} With no weak phases beyond the CKM phase, 
 we also expect
\be\label{R7}
(\sin 2\beta)_{\pi\nu\bar\nu}=(\sin 2\beta)_{\psi K_S}, 
\qquad
(\sin 2\beta)_{\phi K_S}\approx (\sin 2\beta)_{\psi K_S},
\ee
with the accuracy of the last relation at the level of a few percent 
\cite{Worah}.
The confirmation of these two relations would be a very important test of the 
MFV idea. Indeed, in $K\to\pi\nu\bar\nu$ the phase $\beta$ originates in 
the $Z^0$ penguin diagram, whereas in the case of $a_{\psi K_S}$ in 
the $B^0_d-\bar B^0_d$ box diagrams. In the case of the asymmetry 
$a_{\phi K_S}$ it originates also in $B^0_d-\bar B^0_d$ box diagrams, 
but the second relation in (\ref{R7}) could be spoiled by NP 
contributions in the decay amplitude for $B\to \phi K_S$ that is
non-vanishing only at the one loop level.  
Interestingly, the present data from Belle may indicate the
violation of the second relation in (\ref{R7}), although the experimental 
situation is very unclear at present
\cite{browder-talk,Belle-BphiK}:
\begin{equation}\label{aCP-Bd-phiK-mix}
(\sin 2\beta)_{\phi K_S}=
\left\{\begin{array}{ll}
+0.45\pm0.43\pm0.07 &\mbox{(BaBar)}\\
-0.96\pm0.50^{+0.11}_{-0.09} &\mbox{(Belle),}
\end{array}\right.
\end{equation}
A subset of theoretical papers addressing this issue is listed in
\cite{PHIKS}. 

{\bf 5.} An important consequence of (\ref{sin2bnunu})--(\ref{R7}) is the 
following
one. 
For a given $\sin 2\beta$ extracted from $a_{\psi K_S}$ and $Br(\kpn)$ 
only two values of 
$Br(\klpn)$, corresponding to two signs of the master function $X(v)$, 
are possible 
in the full class of MFV models, independent of any new parameters 
present in these models \cite{BF01}. 
Consequently, measuring $Br(\klpn)$ will 
either select one of these two possible values or rule out all MFV models.

{\bf 6.} As pointed out in \cite{BPSW} in most MFV models 
there exists 
a correlation between the zero $\hat s_0$ in the forward-backward asymmetry 
$A_{\rm FB}$ in $B\to X_s\mu^+\mu^-$ and 
$Br(B\to X_s\gamma)$.
We show this correlation in fig.~\ref{corrplot}.

\begin{figure}[hbt]

  \centering 
 \psfragscanon
  \psfrag{bsgammabsgammabsgamma}{ \shortstack{\\ \\ $(Br(B\to
 X_s\gamma)\times 10^4)^\frac12$ ${}$}}
  \psfrag{hats0}[][]{ \shortstack{\\ $\hat{s}_0 $ }}
      \resizebox{.36\paperwidth}{!}{\includegraphics[]{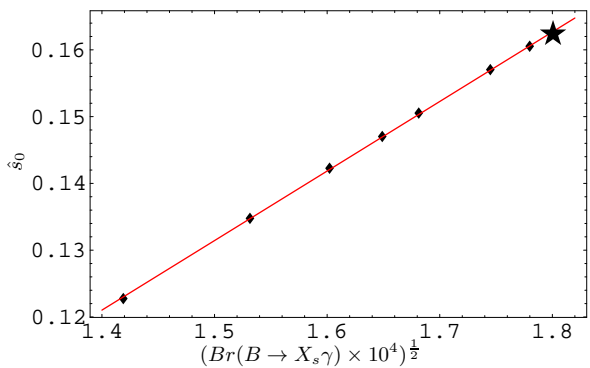}}
    
  \caption[]{\small\label{corrplot} Correlation between
    $\sqrt{Br(B\to X_s\gamma)}$  and $\hat s_0$ \cite{BPSW}. 
    The dots are the results in the ACD
    model (see below) with the compactification scale
    $200,250,300,350,400,600$ and $1000$ GeV  and the star
    denotes the SM value.
}
\end{figure}

{\bf 7.} Other correlations 
between various decays  
can be found in \cite{BF01,Buras:1998ed,Buras:1999da,Bergmann:2001pm,REL}.
For instance there exists in addition to an obvious correlation between 
$K\to\pi\nu\bar\nu$ 
and $B\to X_q\nu\bar\nu$ also a correlation between $\epe$ and rare 
semileptonic $B$ and $K$ decays.
A discussion of correlations between  
$B\to \pi K$ decays and rare decays within MFV models with enhanced 
$Z^0$ penguins can be found in \cite{BFRS-I}.

\subsection{Model Dependent Relations}
\subsubsection{\boldmath{$B_q\to\mu\bar\mu$} and \boldmath{$\Delta M_q$}}
The relations \cite{AJB03}
\be\label{R2}
Br(B_{q}\to\mu\bar\mu)
=4.36\cdot 10^{-10}\frac{\tau(B_{q})}{\hat B_{q}}
\frac{Y^2(v)}{S(v)} 
\Delta M_{q}, \qquad (q=s,d)
\ee
allow to predict $Br(B_{s,d}\to\mu\bar\mu)$  
in a given MFV model, characterized by $Y(v)$ and $S(v)$, with substantially 
smaller hadronic uncertainties 
than found by using directly the formulae for these branching ratios that 
suffer from large uncertainties due to $F_{B_q}$. In particular 
in the SM model we find \cite{AJB03}
\be\label{Results}
 Br(B_{s}\to\mu\bar\mu)=(3.4\pm 0.5)\cdot 10^{-9}, \quad
Br(B_{d}\to\mu\bar\mu)=(1.00\pm 0.14)\cdot 10^{-10},
\ee
where $\overline{m}_t(m_t)=(167\pm 5)\gev$, the lifetimes from \cite{CERNCKM},
 $\hat B_q$ in (\ref{BBB}), $\Delta M_d=(0.503\pm0.006)/{\rm ps}$ 
 and  as an example 
$\Delta M_s=(18.0\pm0.5)/{\rm ps}$, have been used. 
These results
 are substantially more accurate than the ones found in the 
literature in the past but by orders of magnitude below the experimental
upper bounds from CDF(D0) and Belle \cite{Nakao,BelleBd}.

\subsubsection{\bf{$Br(\kpn)$}, \bf{$\Delta M_d/\Delta M_s$ and} 
\bf{$\beta$.} }
In \cite{BB98} an upper bound on $Br(K^+ \rightarrow \pi^+
\nu \bar{\nu})$ in terms of  $\Delta M_d/\Delta M_s$
has been derived within the SM. It has been subsequently cast into 
a useful relation between $Br(\kpn)$, $\Delta M_d/\Delta M_s$ and
$\beta$ in \cite{AI01}. In any MFV model this relation reads
\be \label{AIACD1}
Br(K^+ \rightarrow \pi^+ \nu \bar{\nu})=
7.54\cdot 10^{-6} \vcb^4 X_{\rm eff}^2(v)
\ee
\be \label{AIACD2}
X_{\rm eff}^2(v)= X^2(v)
\Bigg[ \sigma   R^2_t\sin^2\beta+
\frac{1}{\sigma}\left(R_t\cos\beta +
\frac{\lambda^4P_c(X)}{\vcb^2X(v)}\right)^2\Bigg],
\ee
where $\sigma=1/(1-\lambda^2/2)^2$, $\lambda=0.224$,  
$P_c(X)=0.39\pm 0.06$ and $R_t$ is given in (\ref{RTVUB}).
This formula is theoretically  rather clean and does not involve 
hadronic uncertainties except for $\xi$ in (\ref{RTVUB}) and to a lesser 
extent in $\vcb$. 

\subsection{Maximal Enhancements}
How large could various branching ratios in the MFV models be?
A detailed numerical analysis of this
question is beyond the scope of this presentation but, assuming that the 
dominant NP
effects in rare $K$ and $B$ decays come from enhanced $Z^0$ penguins hidden
in the master functions $X(v)$, $Y(v)$ and $Z(v)$, 
bounding this enhancement by the Belle and BaBar data on $B\to X_s l^+l^-$ 
\cite{Kaneko:2002mr}
and setting all other parameters at their central values, we find the results
in column MFV of table~\ref{brMFV} where also 
the SM results are shown.  
While somewhat higher values of branching ratios can still be obtained when 
the input parameters are varied, this exercise shows that enhancements 
of branching ratios in the MFV models by more than factors of six relative 
to the
SM 
should not be expected. 
A similar analysis in a different spirit and with a different set of input 
parameters can be found in 
\cite{AMGIISST}.
\begin{table*}[hbt]
\vspace{0.4cm}
\begin{center}
\caption[]{\small \label{brMFV} Example of branching ratios for rare decays 
in the MFV and  the SM.
}
\begin{tabular}{|c||c|c|}
\hline
{Branching Ratios} &  MFV &   SM
 \\ \hline
$Br(\kpn)\times 10^{11}$ &  $19.1$ &  $8.0$ 
\\ \hline
$Br(\klpn)\times 10^{11}$ &  $9.9$ &   $3.2$ 
\\ \hline
$Br(\kmm)_{\rm SD}\times 10^{9} $ &  $3.5$ &  $0.9$
\\ \hline
$Br(K_{\rm L} \to \pi^0 e^+ e^-)_{\rm CPV}\times 10^{11}$ &  $4.9$ & $3.2$ 
\\ \hline
$Br(B\to X_s\nu\bar\nu)\times 10^{5}$ &  $11.1$ & $3.6$ 
\\ \hline
$Br(B\to X_d\nu\bar\nu)\times 10^{6}$ &  $4.9$ & $1.6$ 
\\ \hline
$Br(B_s\to \mu^+\mu^-)\times 10^{9}$ &  $19.4$ &  $3.9$ 
\\ \hline
$Br(B_d\to \mu^+\mu^-)\times 10^{10}$ &  $6.1$ & $1.2$ 
\\ \hline
\end{tabular}
\end{center}
\end{table*}
\subsection{MFV and Universal Extra Dimensions}
A detailed analysis of all rare and radiative $K$ and $B$ decays and of 
$\Delta M_{d,s}$ in a
particular model with one universal extra dimension (ACD) \cite{ACD} 
has been presented 
in 
\cite{BPSW}. 
The nice feature of this extension of the SM is the presence of only one 
additional parameter, the compactification scale. This feature allows a
unique pattern of various
enhancements and suppressions relative to the SM expectations.
Our analysis shows that all the present 
data on FCNC 
processes are consistent with the compactification scale as low as $300\gev$,
implying that 
the Kaluza--Klein particles could in principle be found already at the 
Tevatron.
Possibly, the most interesting results of our analysis is  
the sizable downward 
shift of  the zero ($\hat s_0$) in the 
$A_{\rm FB}$ asymmetry in $B\to X_s \mu^+\mu^-$ and the suppression of 
$Br(B\to X_s\gamma)$ (also found in \cite{Agashe:2001xt}) that are 
correlated as shown in fig.~\ref{corrplot}. 
Note that a measurement of $\hat s_0$ that is 
higher than the SM estimate would automatically exclude this model as 
there is no compactification scale for which this could be satisfied.
The impact of the Kaluza-Klein particles on $\Delta M_{s,d}$ and on 
the electroweak precision observables can also be found in 
\cite{Chakraverty:2002qk} and \cite{ACD,Papa}, respectively.

\boldmath
\section{Class B: MSSM at Large $\tan\beta$}
\unboldmath
An example of a model in this class is the MSSM with large $\tan\beta$ but
without any relevant contributions from new flavour violating interactions
coming from the non-diagonal elements in squark mass matrices when these are 
given in the quark mass eigenstate basis. In these models the dominant new 
effects in $K$
and $B$ decays come from very strongly enhanced flavour changing neutral 
(FCNC)
Higgs couplings to the down quarks. These couplings are generated only at one
loop level but being proportional to $(\tan\beta)^2$ become very
important for $\tan\beta \ge 30$.

The presence of the enhanced FCNC Higgs couplings implies in turn important
contributions of new operators in the effective theory that are strongly
suppressed in models of class A. In particular the operators
\be
O_S=m_b(\overline{b_R}s_L)(\bar \mu\mu),\qquad 
O_P=m_b(\overline{b_R}s_L)(\bar \mu\gamma_5 \mu) 
\ee
fully dominate the branching ratios for $B_{s,d}\to \mu^+\mu^-$
\cite{BMUMULT} when
$\tan\beta\ge 30$. An approximate formula for $Br(B_{s}\to \mu^+\mu^-)$ is 
given then by \cite{IR02,BCRS}
\be\label{bsmumuH}
Br(B_{s}\to \mu^+\mu^-)\approx 3.5\cdot 10^{-5}
\left[\frac{\tan\beta}{50}\right]^6
\left[\frac{m_t}{M_A}\right]^4 F(\varepsilon_i,\tan\beta)~,
\ee
\be\label{FF}
F(\varepsilon_i,\tan\beta)=
\left[\frac{16\pi^2\varepsilon_Y}
{(1+\tilde\varepsilon_3\tan\beta)(1+\tilde\varepsilon_0\tan\beta)}\right]^2,
\ee
where $\varepsilon_i$, depending on the SUSY parameters, are at most 
$\ord(10^{-2})$ and we have set $\tau(B_s)$, $F_{B_s}$ and $\vts$ at their 
central values. $M_A$ is the mass of the pseudoscalar $A$. 
The expression for $Br(B_{d}\to \mu^+\mu^-)$ is obtained by using
\be\label{ratio}
\frac{Br(B_{d}\to \mu^+\mu^-)}{Br(B_{s}\to \mu^+\mu^-)}=
\left[\frac{\tau(B_d)}{\tau(B_s)}\right]
\left[\frac{F_{B_d}}{F_{B_d}}\right]^2
\left|\frac{V_{td}}{V_{ts}}\right|^2
\left[\frac{M_{B_d}}{M_{B_d}}\right]^5,
\ee
which differs slightly from the usual MFV formula in that the last factor 
has the power five instead of two.
On the other hand the branching ratios themselves can still be
enhanced by almost a factor of 500. If this is indeed the case,
$B_{s,d}\to\mu^+\mu^-$ should be observed already at
Tevatron and $B$ factories, respectively.

The enhanced neutral Higgs couplings and more generally large $\tan\beta$
effects can play also a significant role in 
$\Delta M_s$ \cite{BCRS} inducing  in particular four--fermion operators 
\be
Q^{LR}_1=(\overline{b_L}\gamma_\mu s_L)(\overline{b_R}\gamma^\mu s_R), \qquad
Q^{LR}_2=(\overline{b_R}s_L)(\overline{b_L}s_R),
\ee 
whose Wilson coefficients  are negligible in the SM. 
One finds then
\be\label{DMST}
\Delta M_s=(\Delta M_s)_{\rm SM}(1+f_s)\approx 
(\Delta M_s)_{\rm SM} -|\Delta M_s|^{\rm DP},
\ee
where the new contributions come dominantly from
double Higgs penguins (DP), as indicated above. Being proportional 
to $m_q m_b \tan^4\beta$ for $\Delta M_q$, these contributions can be 
neglected in $\Delta M_d$ that for large $\tan\beta$ is very close to 
the SM estimate. It turns out that   
$\Delta M_s$ is suppressed for any choice of
supersymmetric parameters ($f_s<0$) with the size of suppression dependent 
strongly on the stop mixing, $M_A$ and $\tan\beta$.

As a consequence of the mismatch between Higgs contributions to $\Delta M_d$
and $\Delta M_s$, the MFV formula for $R_t$ in (\ref{RTVUB}) is modified to
\be\label{RT}
R_t=0.90~\left[\frac{\xi}{1.24}\right] \sqrt{\frac{18.4/ps}{\Delta M_s}} 
\sqrt{\frac{\Delta M_d}{0.50/ps}} \sqrt{1+f_s}.
\ee

However, most interesting 
is the correlation of the enhanced neutral Higgs
effects in $B_{s,d}\to \mu^+\mu^-$ and $\Delta M_s$, that is independent of 
$F(\varepsilon_i,\tan\beta)$ in (\ref{FF}):
\be\label{corrH}
Br(B_{s}\to \mu^+\mu^-)\approx 10^{-6}
\left[\frac{\tan\beta}{50}\right]^2
\left[\frac{200\gev}{M_A}\right]^2
\left[\frac{|\Delta M_s|^{\rm DP}}{2.1 /{\rm ps}}\right].
\ee
Consequently a strong enhancement of $B_{s,d}\to \mu^+\mu^-$ implies a
significant suppression of $\Delta M_s$. This means that, in this scenario, an
observation of  $B_{s,d}\to \mu^+\mu^-$ at the level of $\ord(10^{-7})$ and 
$\ord(10^{-8})$ respectively should be accompanied by $\Delta M_s$ below the
SM estimates. 
On the other hand, if $(\Delta M_s)_{\rm exp}>(\Delta M_s)_{\rm
SM}$ is found, this scenario will be  excluded and the observation of
$B_{s,d}\to \mu^+\mu^-$ at this level would point toward other flavour
violating sources, coming for instance from  non-diagonal elements in the
squark mass matrices \cite{CRDP,Dedes}.

The difficult task in testing this scenario will be to demonstrate whether
the measured value $(\Delta M_s)_{\rm exp}$ is indeed smaller or larger than 
$(\Delta M_s)_{\rm SM}$. To this end a significant reduction of the
uncertainties in the non-perturbative parameters is required.
On the other hand an enhancement of $B_{s,d}\to \mu^+\mu^-$ by one or two
orders of magnitude with respect to the SM estimates in (\ref{Results}) would 
be truely spectacular independently of the situation concerning $\Delta M_s$.

Other interesting correlations are the ones between the ratios 
$Br(B\to H\mu^+\mu^-)/Br(B\to H e^+ e^-)$ $(H=K^{(*)},X_s)$ and 
$Br(B_s\to \mu^+\mu^-)$ \cite{HK04}.

\section{Class C: New Weak Phases}
\subsection{Preliminaries}
In this class of models the dominant operators are as in class A but the
master functions become now complex quantities. If the new weak phases
are large, the deviations from the SM can be spectacular, as we
will see below.

\boldmath
\subsection{Weak Phases in $\Delta F=2$ Transitions}
\unboldmath
In the MFV scenario of Section 3 the NP effects enter universally
in $K^0-\bar K^0$,  $B_d^0-\bar B_d^0$ and $B_s^0-\bar B_s^0$ mixing 
through the
single real function $S(v)$, implying strong correlations between new physics
effect in the $\Delta F=2$ observables
of 
$K$ and $B$ decays. When new complex weak phases are present, the situation
could be more involved, with $S(v)$ replaced  by
\be
S_K(v)=|S_K(v)|e^{i\theta_K}, \quad 
S_d(v)=|S_K(v)|e^{i\theta_d}, \quad
S_s(v)=|S_K(v)|e^{i\theta_s},
\ee
for $K^0-\bar K^0$,  $B_d^0-\bar B_d^0$ and $B_s^0-\bar B_s^0$ mixing,
respectively. If these three functions are different from each other, 
some universal properties found in class A are lost. In addition the mixing
induced CP asymmetries in $B$ decays will not measure the angles of the UT
but only sums of these angles and of $\theta_i$. Yet, within each class of 
$K$, $B_d$ and $B_s$ decays, the NP effects of this sort will be
universal. 
Scenarios of this type have been considered for instance in \cite{WPDELTAF2}.
\boldmath
\subsection{Weak Phases in $\Delta F=1$ Transitions}
\unboldmath
New weak phases could enter also decay amplitudes. As now these effects enter
in principle differently in each decay, the situation can be very involved
with many free parameters, no universal effects and little predictive power.

Here I will only discuss one scenario, discussed first in 
\cite{Buras:1998ed,Buras:1999da,BRS,Buchalla:2000sk}
and recently in the context of a simultaneous analysis of prominent non--leptonic
$B$ decays like $B\to\pi\pi$, $B\to\pi K$, $B\to \psi K_S$ and $B\to\phi K_S$
and equally prominent rare decays like $K\to\pi\nu\bar\nu$,
$K_L\to\pi^0e^+e^-$, $B_{s,d}\to\mu^+\mu^-$, $B\to X_{s,d} e^+e^-$ and $\epe$
in \cite{BFRS-III,BFRS-II}. 
It is the scenario of enhanced $Z^0$ penguins with a
large complex weak 
phase
in which the only modification with respect to class A is the replacement 
in the $Z^0$ penguin function
$C(v)\to |C(v)|e^{i\theta_C}$ that makes the master functions $X(v)$, $Y(v)$
and $Z(v)$ complex quantities:
\be\label{THETAS}
X(v)=|X(v)|e^{i\theta_X}, \qquad Y(v)=|Y(v)|e^{i\theta_Y}, \qquad 
Z(v)=|Z(v)|e^{i\theta_Z}.
\ee
The magnitudes and phases of these three functions are correlated with each
other as they depend only on $|C(v)|e^{i\theta_C}$ and other smaller
contributions, that can be set to their SM values.

This new analysis has been motivated
by the experimental situation in $B\to\pi\pi$ and $B\to\pi K$
decays described below. 
While our analysis does not rely on a particular model with the properties
specified above, concrete models with enhanced CP-violating $Z^0$-mediated 
FCNC couplings generated either at the one-loop level or even at the 
tree level have been discussed in the literature. They are reviewed
in \cite{Buras:1998ed,Buras:1999da,BRS,Buchalla:2000sk}, in particular in 
the last of
these papers; see also \cite{GNR97}. Also models with $Z^\prime$-mediated 
FCNCs could be put in this class, provided their contributions can 
effectively be absorbed in the function $C(v)$. For a recent analysis, 
see~\cite{BCLL03}.

\boldmath
\subsection{The $B\to\pi\pi$ Puzzle}
\unboldmath
The BaBar and Belle collaborations have very recently reported the observation 
of $B_d\to\pi^0\pi^0$ decays with CP-averaged branching ratios of 
$(2.1\pm0.6\pm0.3)\times10^{-6}$ and $(1.7\pm0.6\pm0.2)\times10^{-6}$, 
respectively \cite{Babar-Bpi0pi0,Belle-Bpi0pi0}. These measurements represent
quite some challenge for theory. For example, in a recent state-of-the-art 
calculation \cite{Be-Ne} within QCD factorization (QCDF) \cite{BBNS1}, 
a branching 
ratio that is about six times smaller is favoured, whereas the calculation of
$B_d\to\pi^+\pi^-$ points towards a branching ratio about two times 
larger than the current experimental average. On the other hand, the
calculation of $B^+\to\pi^+\pi^0$ reproduces the data rather well. This
``$B\to\pi\pi$ puzzle'' is reflected by the following quantities:
\begin{eqnarray}
R_{+-}^{\pi\pi}&\equiv&2\left[\frac{Br(B^\pm\to\pi^\pm\pi^0)}{Br
(B_d\to\pi^+\pi^-)}\right]\frac{\tau_{B^0_d}}{\tau_{B^+}}=
2.12\pm0.37~~\mbox{}\label{Rpm-def}\\
R_{00}^{\pi\pi}&\equiv&2\left[\frac{Br(B_d\to\pi^0\pi^0)}{Br
(B_d\to\pi^+\pi^-)}\right]=0.83\pm0.23.\label{R00-def}
\end{eqnarray}
The central values calculated in QCDF 
\cite{Be-Ne} are $R_{+-}^{\pi\pi}=1.24$ and $R_{00}^{\pi\pi}=0.07$.
\boldmath
\subsection{The $B\to\pi K$ Puzzle}
\unboldmath
In the $B\to\pi K$ system, the CLEO, BaBar and Belle collaborations have 
measured the following ratios of 
CP-averaged branching ratios \cite{BF-neutral1}:
\begin{eqnarray}
R_{\rm c}&\equiv&2\left[\frac{Br(B^\pm\to\pi^0K^\pm)}{Br
(B^\pm\to\pi^\pm K^0)}\right]=1.17\pm0.12\label{Rc-def}\\
R_{\rm n}&\equiv&\frac{1}{2}\left[
\frac{Br(B_d\to\pi^\mp K^\pm)}{Br(B_d\to\pi^0K)}\right]=
0.76\pm0.10,\label{Rn-def}
\end{eqnarray}
with numerical values following from \cite{HFAG}. As noted in 
\cite{BF-neutral2}, the pattern of $R_{\rm c}>1$ and $R_{\rm n}<1$, which
is now consistently favoured by the separtate BaBar, Belle and CLEO data, 
is actually  puzzling in the framework of QCD factorization \cite{Be-Ne}
 that gives 
typically $R_{\rm c}\approx R_{\rm n}\approx 1.15$. 
This is clearly seen in the  $(R_{\rm n}, R_{\rm c})$ plot in 
fig.~\ref{fig:Rn-Rc}
\cite{BFRS-III}, to which we will return below.
On the other hand,
\begin{equation}\label{R-def}
R\equiv\left[\frac{Br(B_d\to\pi^\mp K^\pm)}{Br
(B^\pm\to\pi^\pm K)}\right]\frac{\tau_{B^+}}{\tau_{B^0_d}}
=0.91\pm0.07
\end{equation}
does not show any anomalous behaviour. Since $R_{\rm c}$ and $R_{\rm n}$ are 
affected significantly by colour-allowed EW penguins, whereas this is not the
case of $R$, this ``$B\to\pi K$ 
puzzle'' may be a manifestation of NP in the EW penguin sector 
\cite{BFRS-III,BFRS-II,BFRS-I,BF-neutral2}, offering an attractive avenue for 
physics beyond the SM to enter the $B\to\pi K$ system \cite{FM-NP,trojan}.

\begin{figure}
\vspace*{0.3truecm}
\begin{center}
\includegraphics[width=10cm]{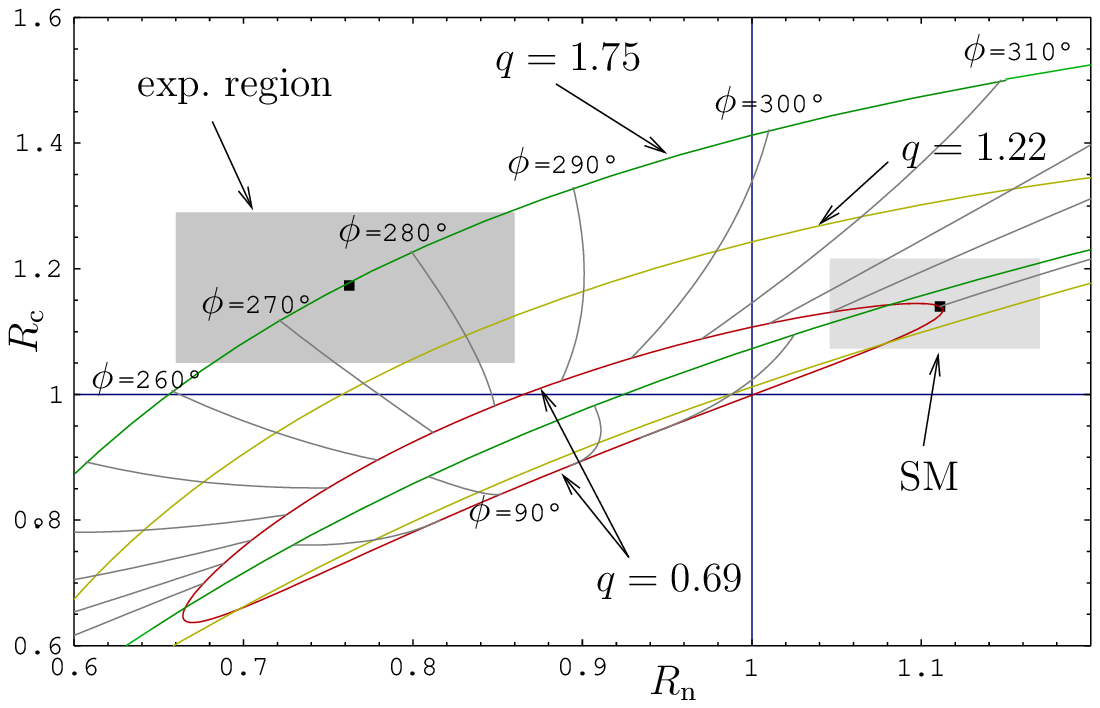}
\end{center}
\caption{The situation in the $R_{\rm n}$--$R_{\rm c}$ plane \cite{BFRS-III}.
We show contours
for values of $q=0.69$, $q=1.22$ and $q=1.75$ and 
$\phi \in [0^\circ,360^\circ]$. The
ranges from (\ref{Rc-def}) and (\ref{Rn-def}) (experiment) as well as 
from the SM are indicated in grey.}\label{fig:Rn-Rc}
\end{figure}

\subsection{The Analysis of \cite{BFRS-III,BFRS-II}}
In view of significant experimental uncertainties, none of these 
exciting results is conclusive at the moment, but it is legitimate and 
interesting to take them seriously and to search for possible origins of 
these ``signals'' for deviations from the SM expectations. As we are 
dealing here with non-leptonic decays, the natural question arises whether 
these signals originate in the NP contributions or/and result from our 
insufficient understanding of hadron dynamics. 
The purpose of \cite{BFRS-III,BFRS-II}
was to develop a strategy which would allow 
us to address the $B\to\pi\pi$ and $B\to\pi K$ puzzles in a systematic manner
once the experimental 
data on these decays improve. In order to illustrate this strategy in 
explicit terms, we considered a simple extension of the SM in which NP enters
dominantly through 
enhanced CP-violating $Z^0$ penguins. As we will see below, this choice is 
dictated by the pattern of the data on the $B\to\pi K$ observables and 
the great predictivity of this scenario. It was first considered in 
\cite{Buras:1998ed,Buras:1999da,BRS} to study correlations between rare $K$ 
decays and the ratio $\epe$, and was generalized to rare $B$ decays in 
\cite{Buchalla:2000sk}. Extending these considerations to non-leptonic
$B$-meson decays, allowed us to confront this NP scenario with
many more experimental results. Our strategy consists of three interrelated 
steps, and has the following logical structure:

\vspace*{0.3truecm}
\newpage
\noindent
{\bf Step 1:}

\noindent
Since $B\to\pi\pi$ decays and the usual analysis of the UT are only 
insignificantly affected by EW penguins, the $B\to\pi\pi$ system can be 
described as in the SM. Employing the $SU(2)$ isospin flavour symmetry
of strong interactions and the information on $\gamma$ from the
UT fits \cite{CERNCKM}, we could extract the relevant hadronic $B\to\pi\pi$ 
parameters, and find large 
non-factorizable contributions, in variance with 
the QCD factorization approach. 
 Having these parameters at hand, we 
could then also predict the direct and mixing-induced CP asymmetries
of the $B_d\to\pi^0\pi^0$ channel. A future measurement of one of these
observables allows a determination of $\gamma$.

\vspace*{0.3truecm}

\noindent
{\bf Step 2:}

\noindent
Using the $SU(3)$ flavour symmetry and plausible dynamical 
assumptions, we could subsequently determine the hadronic $B\to\pi K$ 
parameters 
through the $B\to\pi\pi$ analysis, and  calculate the $B\to\pi K$ 
observables in the SM. Interestingly, we found agreement with the pattern 
of the $B$-factory data for those observables where EW penguins play only 
a minor r\^ole. On the other hand, the observables receiving significant
EW penguin contributions did {\it not} agree with the experimental picture, 
thereby suggesting NP in the EW penguin sector. 
Parametrizing the EW contributions to $B\to\pi K$ by $qe^{i\phi}$ that 
is related to the $Z^0$-penguin function $C(v)$ through \cite{BFRS-I}
\be\label{RG}
C(v)=|C(v)|e^{i\theta_C} = 2.35\, \bar q e^{i\phi} -0.82,\quad 
\bar q= q \left[\frac{|V_{ub}/V_{cb}|}{0.086}\right], 
\ee
we have demonstrated
\cite{BFRS-III,BFRS-II} that one can describe all the currently 
available $B\to\pi K$ data 
provided 
\begin{equation}\label{q-det}
q=1.75^{+1.27}_{-0.99},\quad  \phi=-(85^{+11}_{-14})^\circ,
\end{equation}
to be compared with $q=0.69$ \cite{NR} and $\phi=0$ in the SM. In particular 
as seen in fig.~\ref{fig:Rn-Rc} the values in (\ref{q-det}) allow to 
describe properly the data in
(\ref{Rc-def}) and (\ref{Rn-def}).
The requirement of a large CP-violating NP phase around $-90^\circ$ is the 
most interesting result of this study.
A crucial future test of 
this scenario will be provided by the CP-violating $B_d\to \pi^0 K_{\rm S}$ 
observables, which we could predict. Moreover, we could obtain valuable
insights 
into $SU(3)$-breaking effects, which support our working assumptions, and 
could also determine the UT angle $\gamma$, that is in  
agreement with the UT fits.  

\vspace*{0.3truecm}

\noindent
{\bf Step 3:}

\noindent
In turn, the sizeably enhanced EW penguins with their large CP-violating 
NP phase have important implications for rare $K$ and $B$ decays as well as 
$\epe$.
Interestingly, several predictions differ significantly from the SM 
expectations and should easily be identified once the data improve. 

\vspace*{0.3truecm}

Including the constraint from $B\to X_s l^+l^-$ \cite{Kaneko:2002mr}, 
that selects $q\le 1.0$ 
in (\ref{q-det}), the most interesting results of this study,  
presented in \cite{BFRS-III,BFRS-II} are: 

a) 
For the very clean $K\to\pi\nu\bar\nu$ decays, we find
\begin{equation}\label{kpnr}
\begin{array}{rcl}
Br(\kpn)=(7.5\pm 2.1)\cdot 10^{-11}, \\
Br(\klpn)=(3.1\pm 1.0)\cdot 10^{-10},
\end{array}
\end{equation}
to be compared with the SM predictions, for which we find  
$(8.0 \pm 1.1)\times 10^{-11}$ and 
$(3.2 \pm 0.5)\times 10^{-11}$  in the ballpark of other estimates 
\cite{Gino03},
respectively. On the experimental side the results of 
the AGS E787 \cite{Adler97} 
collaboration and  
 KTeV \cite{KTeV00X}  are 
$Br(K^+ \rightarrow \pi^+ \nu \bar{\nu})=
(15.7^{+17.5}_{-8.2})\times 10^{-11}$ and 
$Br(\klpn)< 5.9\times 10^{-7}$, respectively.
The enhancement of $Br(\klpn)$ by one order of magnitude 
and the pattern in (\ref{kpnr}) are dominantly the 
consequences of $\beta_X=\beta-\theta_X\approx
111^\circ$ with $\theta_X$ defined in (\ref{THETAS}).
Indeed,
\be
\frac{Br(\klpn)}{Br(\klpn)_{\rm SM}}=
\left|\frac{X}{X_{\rm SM}}\right|^2
\left[\frac{\sin\beta_X}{\sin\beta}\right]^2
\ee
\be
\frac{Br(\klpn)}{Br(\kpn)}\approx 4.4\times (\sin\beta_X)^2
\approx (4.2\pm 0.2).  
\ee
Interestingly, the above ratio turns out to be very close to its absolute
upper bound in \cite{GRNIR}. 
Fig.~\ref{fig:KpKl} illustrates nicely these features.
A spectacular 
implication of these findings is a strong violation of 
$(\sin 2 \beta)_{\pi \nu\bar\nu}=(\sin 2 \beta)_{\psi K_{\rm S}}$ 
\cite{BBSIN}, which is valid in the SM and any model with minimal
flavour violation as discussed in Section 3. Indeed, we find
\be
(\sin 2 \beta)_{\pi \nu\bar\nu}=\sin 2\beta_X= -(0.69^{+0.23}_{-0.41}),
\ee
in striking disagreement with $(\sin 2 \beta)_{\psi K_{\rm S}}= 0.74\pm0.05$ 
in (\ref{ga}).
  
b) Another implication is the large branching ratio
\be
Br(K_{\rm L}\to\pi^0e^+e^-)= (7.8\pm 1.6)\times 10^{-11},
\ee
which is governed by direct CP violation in this scenario. 
On the other hand, the
SM result $(3.2^{+1.2}_{-0.8})\times 10^{-11}$ \cite{BI03} is dominated 
by the indirect CP violation. Next, the integrated forward--backward CP 
asymmetry for $B_d\to K^*\mu^+\mu^-$ \cite{Buchalla:2000sk} 
can be very large in view of $\theta_Y\approx -100^\circ$ as
it is given by
\be
A^{\rm CP}_{\rm FB}=(0.03\pm0.01)\times \tan\theta_Y. 
\ee

c)  
Next, $Br(B\to X_{s,d}\nu\bar\nu)$ and 
$Br(B_{s,d}\to \mu^+\mu^-)$ are enhanced by factors of $2$ and $5$, 
respectively. The impact on $K_{\rm L}\to \mu^+\mu^-$ is rather 
moderate.

d) 
As emphasized in \cite{Buras:1998ed}, enhanced $Z^0$ penguins may play an 
important r\^ole in $\epe$. The enhanced value of $C$ and its large negative 
phase suggested by the $B\to\pi K$ analysis require a significant enhancement 
of the relevant hadronic matrix element of the QCD penguin operator $Q_6$, 
with respect to the one of the EW penguin operator $Q_8$, to be 
consistent with the $\epe$ data. 

e) 
We have also explored the implications for the decay $B_d\to\phi K_{\rm S}$ 
\cite{BFRS-III}. Large hadronic uncertainties preclude a precise 
prediction, but assuming that the sign of the cosine 
of a strong phase agrees with factorization, we find that 
$(\sin 2 \beta)_{\phi K_{\rm S}}>(\sin 2 \beta)_{\psi K_{\rm S}}$.
This pattern is qualitatively different from the present $B$-factory 
data \cite{browder-talk}, which are, however, not yet conclusive as seen in 
(\ref{aCP-Bd-phiK-mix}).
On the other hand, a future confirmation of this pattern could be 
another signal of enhanced CP-violating $Z^0$ penguins at work.

\begin{figure}
\vspace*{0.3truecm}
\begin{center}
\includegraphics[width=10cm]{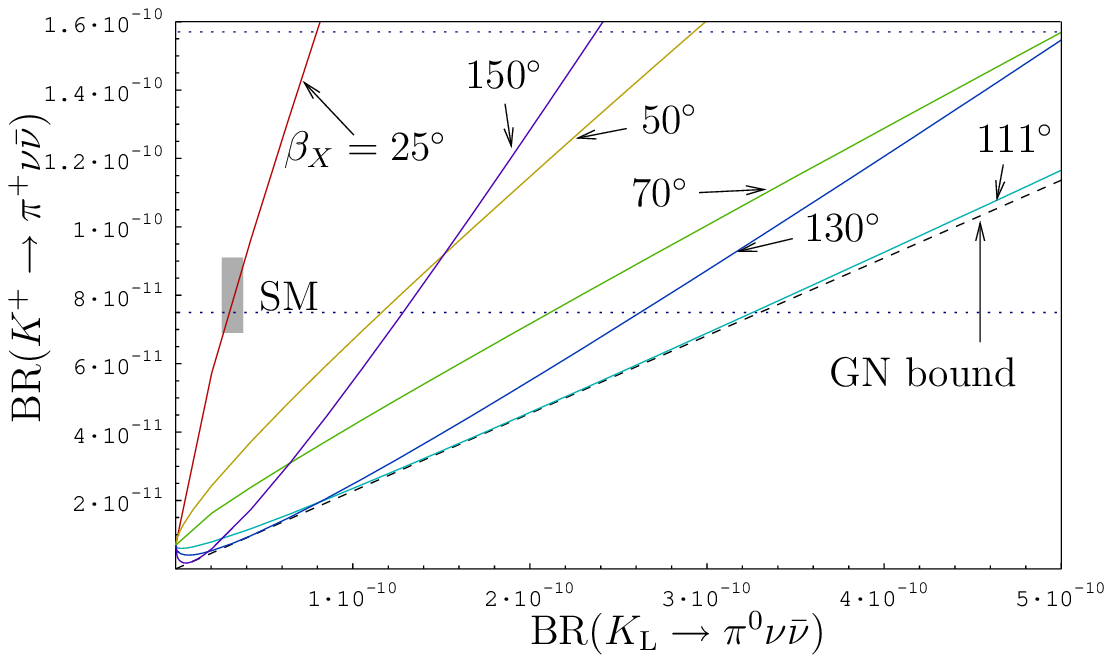}
\end{center}
\caption{${Br}(\kpn)$ as a function of ${Br}(\klpn)$
for various values of $\beta_X$ \cite{BFRS-III}. The dotted horizontal 
lines indicate 
the lower part of the experimental range
 and the grey area the SM prediction. We also show the 
bound of \cite{GRNIR}.  \label{fig:KpKl}}
\end{figure}

\boldmath
\section{Shopping List}
\unboldmath
We have seen that each of the NP scenarios discussed above had some specific
features not shared by other scenarios and with a sufficient number of
measurements it should be possible to distinguish between them, eventually
selecting one of them or demonstrating the necessity for going to scenarios
in classes D and E.

There is a number of questions which I hope will be answered in the coming
years:
\begin{itemize}
\item
Probably the most important at present is the clarification of the
discrepancy between Belle \cite{Belle-BphiK} and BaBar \cite{browder-talk}
 in the measurement of 
$(\sin2\beta)_{\phi K_S}$. The confirmation of the significant departure of 
$(\sin2\beta)_{\phi K_S}$ from the already accurate value of 
$(\sin2\beta)_{\psi K_S}$, would be a clear signal of new physics that 
cannot be accomodated within classes A and B.
Also in a particular scenario of class C, just discussed, one finds 
$(\sin2\beta)_{\phi K_S}>(\sin2\beta)_{\psi K_S}$ and consequently 
$(\sin 2\beta)_{\phi K_{\rm S}}$ is of the same magnitude as the 
central value found by the Belle collaboration but of {\it opposite} 
sign! Thus it is likely that the confirmation of the Belle result will 
require the models of class D.
\item
Also very important are the measurements of $Br(B_{d,s}\to \mu^+\mu^-)$ and
$\Delta M_s$. The possible enhancements of $Br(B_{d,s}\to \mu^+\mu^-)$ by 
 factors as high as $500$, discussed in Section 4, are the largest 
enhancements in the field of $K$
and $B$ decays, that are still consistent with all available data. The
measurement of $\Delta M_s$ is, on the other hand, very important as it may
have a considerable impact on the determination of the unitarity triangle.
Finding $\Delta M_s$ below $(\Delta M_s)_{\rm SM}$ and $Br(B_{d,s}\to
\mu^+\mu^-)$ well above the SM expectations would be a nice confirmation 
of a SUSY scenario with a large $\tan\beta$ that we discussed in Section 4.
\item
The improved measurements of several $B\to\pi\pi$ and $B\to\pi K$ observables
are very important in order to see whether the theoretical approaches like 
QCDF \cite{BBNS1}, PQCD \cite{PQCD} and SCET \cite{SCET} in addition to their
nice theoretical structures are also
phenomenologically useful. On the other hand, independently of the outcome of
these measurements, the pure phenomenological strategy
\cite{BFRS-III,BFRS-II} presented in Section
5, will be useful in correlating the experimental results for $B\to\pi\pi$ and
$B\to\pi K$ with those for rare $K$ and $B$ decays, $B_d\to\phi K_S$ and 
$\epe$.
\item
Assuming that the future more accurate data on $B\to\pi\pi$ and
$B\to\pi K$ will not modify significantly the presently observed pattern in
these decays, the scenario of enhanced $Z^0$ penguins with a new large
complex weak phase will remain to be an attractive possibility. While the 
enhancement of $Br(\klpn)$ by one order of magnitude would be very welcome to
our experimental colleagues and $(\sin2\beta)_{\pi\nu\bar\nu}<0$
 would be a very spectacular signal of NP, 
even more moderate departures of this sort
 from the SM and the MFV expectations could be easily identified in the very
clean $K\to\pi\nu\bar\nu$ decays as clear signals of NP.
\item
The improved measurements of $Br(B\to X_sl^+l^-)$ and $Br(\kpn)$ in the
coming years will efficiently bound the possible enhancements of
$Z^0$ penguins, at least within the scenarios A--C discussed here.
\item
Also very important is an improved measurement of $Br(B\to X_s\gamma)$ as
well as the removal of its sensitivity to $\mu_c$ in $m_c(\mu_c)$ through a
NNLO calculation. This would increase the precision on the MFV correlation   
between $Br(B\to X_s\gamma)$ and the zero $\hat s_0$ in the forward--backward
asymmetry $A_{\rm FB}(\hat s)$ in $B\to X_s l^+l^-$. A $20\%$ suppression of 
$Br(B\to X_s\gamma)$ with respect to the SM accompanied by a downward shift
of $\hat s_0$ would be an interesting confirmation of the correlation in
question and consistent with the effects of universal extra dimensions with 
a low compactification scale of order few hundred GeV. On the other hand 
finding no zero in $A_{\rm FB}(\hat s)$ would
likely point towards flavour violation beyond the MFV.
\item
Finally, improved bounds and/or measurements of processes not existing or
very strongly suppressed in the SM, like various electric dipol moments and
FCNC transitions in the charm sector will be very important in the search for
new physics. The same applies to $\mu\to e\gamma$ and generally lepton
flavour violation.
\end{itemize}

We could continue this list for ever, in particular in view of the
expected progress at Belle and Babar, charm physics at Cornell,
experimental program at LHCb, BeTeV and the planned rare $K$ physics 
experiments.
But the upper bound on the maximal number of pages for my contribution has
been already significantly violated which is a clear signal that I should 
conclude here.
The conclusion is not unexpected:
in this decade, it will be very 
exciting to follow the development in this field  and
to monitor the values of various observables provided by our experimental
colleagues by using the strategies presented here and other strategies 
found in the rich literature.

{\bf Acknowledgements}

I would like to thank the organizers for inviting me to such a 
wonderful meeting  and enjoyable atmosphere, and Felix Schwab for invaluable
comments on the manuscript.
Most importantly, the material presented here would not be possible without
such a fruitful collaboration with 20 magnificant physicists whose names 
can be found below. Many thanks to all of them.
The work presented here has been supported in part by the German 
Bundesministerium f\"ur
Bildung und Forschung under the contract 05HT1WOA3 and the 
DFG Project Bu. 706/1-2.

%


\begin{thebibliography}{80}
\bibitem{Weinberg:2004kv}
S.~Weinberg,
arXiv:hep-ph/0401010.
\bibitem{GIM}
{ S.L. Glashow, J. Iliopoulos and L. Maiani}
{ Phys. Rev.} {\bf D2},  1285 (1970).
\bibitem{CKM}
N. Cabibbo, Phys. Rev. Lett. {\bf 10},  531 (1963);
M. Kobayashi and T. Maskawa,
{ Prog. Theor. Phys.}~{\bf 49}  652 (1973).

\bibitem{REV}
R. Fleischer,
{Phys.\ Rep.}~{\bf 370}  537 (2002).
Y. Nir, hep-ph/0109090, G. Buchalla,  
hep-ph/0302145;
A.~J.~Buras,
arXiv:hep-ph/0307203.
A.~Ali,
arXiv:hep-ph/0312303;
T.~Hurth,
Rev.\ Mod.\ Phys.\  {\bf 75}, 1159 (2003).
G.~Isidori,
hep-ph/0401079.






\bibitem{AJBLH}
A.~J.~Buras,
arXiv:hep-ph/9806471.









\bibitem{BBL}
G. Buchalla,
 A.J. Buras and M.E. Lautenbacher,
{ Rev. Mod. Phys} {\bf 68},  1125 (1996).
\bibitem{Pisa}
A.~J.~Buras,
arXiv:hep-ph/0109197.







\bibitem{IL}
{ T. Inami and C.S. Lim,}
{ Progr. Theor. Phys.} {\bf 65},  297 (1981).

\bibitem{Erice}
A.~J.~Buras,
arXiv:hep-ph/0101336.







\bibitem{UUT}
A.~J.~Buras, P.~Gambino, M.~Gorbahn, S.~J\"ager and L.~Silvestrini,
Phys.\ Lett.\ B {\bf 500}, 161 (2001).






\bibitem{AMGIISST} 
G.~D'Ambrosio,
 G.~F.~Giudice, G.~Isidori and A.~Strumia,
Nucl.\ Phys.\ B {\bf 645}, 155 (2002).


\bibitem{BOEWKRUR} 
C.~Bobeth, T.~Ewerth, F.~Kruger and J.~Urban,
Phys.\ Rev.\ D {\bf 66}, 074021 (2002).




\bibitem{Cracow}
A.J. Buras,
[hep-ph/0310208].

\bibitem{MPR}
M. Misiak, S. Pokorski and J. Rosiek, hep-ph/9703442, 

\bibitem{GGMS} 
 F. Gabbiani,
 E. Gabrielli, A. Masiero and L. Silvestrini,
 { Nucl. Phys.} {\bf B477}  321 (1996). 
M.~Ciuchini,
 E.~Franco, A.~Masiero and L.~Silvestrini,
Phys.\ Rev.\ D {\bf 67}, 075016 (2003)
[Erratum-ibid.\ D {\bf 68}, 079901 (2003)].




\bibitem{UNIT}
H.~Abele { et al.},
arXiv:hep-ph/0312150.


\bibitem{CERNCKM}
M.~Battaglia, A.J. Buras, P. Gambino, A. Stocchi {\it et al.},
hep-ph/0304132.



\bibitem{BaBar}
B. Aubert et al., Phys. Rev. Lett. {\bf 89},
 201802 (2002).
\bibitem{Belle}
K. Abe et al., Phys. Rev. {\bf D66},  071102 (2002).









\bibitem{BUPAST} 
A.~J.~Buras, F.~Parodi and A.~Stocchi,
JHEP {\bf 0301}, 029 (2003).



\bibitem{AJB03}
A.~J.~Buras,
Phys.\ Lett.\ B {\bf 566}, 115 (2003).


\bibitem{BBSIN}
G. Buchalla and  A.~J.~Buras,
{ Phys.\ Lett.}~{\bf B333}, 221 (1994).
Phys.\ Rev.\ D {\bf 54}, 6782 (1996).



\bibitem{BFRS-III}
A.~J.~Buras, R.~Fleischer, S.~Recksiegel and F.~Schwab,
arXiv:hep-ph/0402112.



\bibitem{BFRS-II}
A.~J.~Buras, R.~Fleischer, S.~Recksiegel and F.~Schwab,
to appear in {\it Phys.\ Rev. Lett.}, 
arXiv:hep-ph/0312259.




\bibitem{BF01}
A.~J.~Buras and R.~Fleischer,
Phys.\ Rev.\ D {\bf 64}, 115010 (2001).



\bibitem{Worah}
Y. Grossman, G. Isidori and M.P. Worah, Phys. Rev. {\bf D58},  057504
(1998).


\bibitem{browder-talk}T.E. Browder, 
hep-ex/0312024.


\bibitem{Belle-BphiK}K. Abe {\it et al.}\ [Belle Collaboration],
hep-ex/0308035.



\bibitem{PHIKS}
R. Fleischer and T. Mannel, { Phys. Lett.} {\bf B511},  240 (2001),
G. Hiller, Phys. Rev. {\bf D66},  071502 (2002),
A. Datta, Phys. Rev. {\bf D66},  071702 (2002),
M. Ciuchini and L. Silvestrini, {Phys. Rev. Lett.} {\bf 89},  231802 
(2002),
M. Raidal, {Phys. Rev. Lett.} {\bf 89},  231803 (2002),
Y. Grossman, Z. Ligeti, Y. Nir and H. Quinn, hep-ph/0303171,
S. Khalil and E. Kou, hep-ph/0307024.





\bibitem{BPSW}
A.~J.~Buras, M.~Spranger and A.~Weiler,
Nucl.\ Phys.\ B {\bf 660}, 225 (2003);
A.~J.~Buras, A.~Poschenrieder, M.~Spranger and A.~Weiler,
Nucl.\ Phys.\ B {\bf 678}, 455 (2004).




\bibitem{Buras:1998ed}
A.~J.~Buras and L.~Silvestrini,
Nucl.\ Phys.\ B {\bf 546}, 299 (1999).

\bibitem{Buras:1999da}
A.~J.~Buras, G.~Colangelo, G.~Isidori, A.~Romanino and L.~Silvestrini,
Nucl.\ Phys.\ B {\bf 566}, 3 (2000).

\bibitem{Bergmann:2001pm}
S.~Bergmann and G.~Perez,
Phys.\ Rev.\ D {\bf 64}, 115009 (2001),
%
JHEP {\bf 0008}, 034 (2000).



\bibitem{REL} S. Laplace, Z. Ligeti, Y. Nir and G. Perez, 
{ Phys. Rev.} {\bf D65},  094040 (2002); 
G.~Buchalla and A.~S.~Safir,
arXiv:hep-ph/0310218.




\bibitem{BFRS-I}
A.~J.~Buras, R.~Fleischer, S.~Recksiegel and F.~Schwab,
Eur.\ Phys.\ J.\ C {\bf 32}, 45 (2003).






\bibitem{Nakao}
M.~Nakao,
arXiv:hep-ex/0312041.



\bibitem{BelleBd}
M.-C. Chang et al, Belle Collaboration, hep-ex/0309069.


\bibitem{BB98}
G.~Buchalla and A.~J.~Buras,
Nucl.\ Phys.\ B {\bf 548}, 309 (1999).



\bibitem{AI01}
G. D'Ambrosio and G. Isidori, 
{ Phys. Lett.} {\bf B530},  108 (2002).



\bibitem{Kaneko:2002mr}
J.~Kaneko {\it et al.}  [Belle Collaboration],
Phys.\ Rev.\ Lett.\  {\bf 90},  021801 (2003),
B.~Aubert {\it et al.}  [BABAR Collaboration],
arXiv:hep-ex/0308016.

\bibitem{ACD}
T.~Appelquist, 
H.~C.~Cheng and B.~A.~Dobrescu,
Phys.\ Rev.\ D {\bf 64}, 035002 (2001).




\bibitem{Agashe:2001xt} 
K.~Agashe, N.~G.~Deshpande and G.~H.~Wu,
Phys.\ Lett.\ B {\bf 514},  309 (2001).


\bibitem{Chakraverty:2002qk}
D.~Chakraverty, K.~Huitu and A.~Kundu,
Phys.\ Lett.\ B {\bf 558}, 173 (2003).


\bibitem{Papa}
J.~Papavassiliou and A.~Santamaria,
Phys.\ Rev.\ D {\bf 63}, 016002 (2001),
J.~F.~Oliver, J.~Papavassiliou and A.~Santamaria,
Phys.\ Rev.\ D {\bf 67}, 056002 (2003).





\bibitem{BMUMULT}
C.-S.~Huang and Q.-S.~Yan, Phys. Lett. B {\bf 442}  209 (1998);
C.-S.~Huang,
 W.~Liao and Q.-S.~Yan, 
Phys. Rev. D {\bf 59}  011701 (1999);
K.~S.~Babu and C.~Kolda, Phys. Rev. Lett. {\bf 84}  228 (2000);
P.~H.~Chankowski and L.~Slawianowska, Phys. Rev. D {\bf 63}  054012 (2001);
C.~S.~Huang, 
 W.~Liao, Q.~S.~Yan and S.~H.~Zhu, 
Phys. Rev.  D {\bf 63}  114021 (2001);
Phys. Rev. {\bf  64}  059902 (2001) (E);
C.~Bobeth,
 T.~Ewerth, F.~Kr\"uger and J.~Urban, 
Phys. Rev. D {\bf 64} 074014 (2001).


\bibitem{IR02}
G.~Isidori and A.~Retico,
JHEP {\bf 0111}, 001 (2001).



\bibitem{BCRS}
A.~J.~Buras, P.~H.~Chankowski, J.~Rosiek and L.~Slawianowska,
Nucl.\ Phys.\ B {\bf 619}, 434 (2001);
Phys.\ Lett.\ B {\bf 546}, 96 (2002);
Nucl.\ Phys.\ B {\bf 659}, 3 (2003).





\bibitem{CRDP}
P.~H.~Chankowski and J.~Rosiek,
Acta Phys.\ Polon.\ B {\bf 33}, 2329 (2002).
G.~Isidori and A.~Retico,
JHEP {\bf 0209}, 063 (2002).

\bibitem{Dedes}
A.~Dedes and A.~Pilaftsis,
Phys.\ Rev.\ D {\bf 67}, 015012 (2003),
A.~Dedes,
Mod.\ Phys.\ Lett.\ A {\bf 18}, 2627 (2003).








\bibitem{HK04}
G.~Hiller and F.~Kruger,
arXiv:hep-ph/0310219.



\bibitem{WPDELTAF2}
S.~Bertolini, F.~Borzumati and A.~Masiero,
Phys.\ Lett.\ B {\bf 194}, 545 (1987)
[Erratum-ibid.\ B {\bf 198}, 590 (1987)].
Y.~Nir and D.~J.~Silverman,
Nucl.\ Phys.\ B {\bf 345}, 301 (1990);
Phys.\ Rev.\ D {\bf 42}, 1477 (1990);
R. Fleischer, G. Isidori and J. Matias, 
{ JHEP} {\bf 0305}, 053 (2003).






\bibitem{BRS}
A.J. Buras, A.~Romanino and L.~Silvestrini,
{ Nucl.\ Phys.}~{\bf B520}, 3 (1998).





\bibitem{Buchalla:2000sk}
G.~Buchalla, G.~Hiller and G.~Isidori,
Phys.\ Rev.\ D {\bf 63}, 014015 (2001),
D. Atwood and G. Hiller,
[hep-ph/0307251]. 



\bibitem{GNR97}
Y.~Grossman, Y.~Nir and R.~Rattazzi,
{\it Adv.\ Ser.\ Direct.\ High Energy Phys.}~{\bf 15} (1998) 755.


\bibitem{BCLL03}
V. Barger, C.W. Chiang, P. Langacker and H.S. Lee,
{\it Phys.\ Lett.}~{\bf B580} (2004) 186.






\bibitem{Babar-Bpi0pi0}B. Aubert {\it et al.}\  [BaBar Collaboration],
[hep-ex/0308012].

\bibitem{Belle-Bpi0pi0}K. Abe {\it et al.}\  [Belle Collaboration],
[hep-ex/0308040].

\bibitem{Be-Ne}
M. Beneke and M. Neubert,
{ Nucl.\ Phys.}~{\bf B675}  333 (2003).






\bibitem{BBNS1}M. Beneke,
 G. Buchalla, M. Neubert and C.T. Sachrajda,
{ Phys.\ Rev.\ Lett.}~{\bf 83}  1914 (1999).







\bibitem{BF-neutral1} A.J. Buras and R. Fleischer,
{ Eur.\ Phys.\ J.}~{\bf C11}  93 (1999).



\bibitem{HFAG}Heavy Flavour Averaging Group,
{\tt http://www.slac.stanford.edu/xorg/hfag/}.





\bibitem{BF-neutral2}A.J. Buras and R. Fleischer,
{ Eur.\ Phys.\ J.}~{\bf C16}  97 (2000).










\bibitem{FM-NP}R. Fleischer and T. Mannel,
 [hep-ph/9706261].


\bibitem{trojan}Y. Grossman, M. Neubert and A. Kagan,
{ JHEP} {\bf 9910}, 029 (1999);
{ Phys.\ Rev.}~{\bf D60}, 034021 (1999).
T.~Yoshikawa,
{ Phys.\ Rev.}~{\bf D68}, 054023 (2003).
M.~Gronau and J.~L.~Rosner,
arXiv:hep-ph/0311280,
{\it Phys.\ Lett.}~{\bf B572} (2003) 43.
















\bibitem{NR}M. Neubert and J.L. Rosner,
{\it Phys.\ Lett.}~{\bf B441}  403 (1998);
{\it Phys.\ Rev.\ Lett.}~{\bf 81}  5076 (1998).




\bibitem{Gino03}
G. Isidori,
hep-ph/0307014 
 and references therein.




\bibitem{Adler97}
S. Adler {\it et al.},
  [E787 Collaboration],
{ Phys.\ Rev.\ Lett.}~{\bf 88}, 041803 (2002).


\bibitem{KTeV00X}
A. Alavi-Harati {\it et al.}, {\it Phys.\ Rev.}~{\bf D61}  072006 (2000).




\bibitem{GRNIR}
Y. Grossman and Y. Nir,
{ Phys.\ Lett.}~{\bf B398}, 163 (1997).



\bibitem{BI03}
G.~Buchalla, G.~D'Ambrosio and G.~Isidori,
arXiv:hep-ph/0308008.
%



\bibitem{PQCD}
Y.-Y. Keum {\it et al.} 
{ Phys. Rev.} {\bf D63},  054008 (2001) and references therein.

\bibitem{SCET}
Ch.W. Bauer et al.,  { Phys. Rev. }  {\bf D65},  054022 (2002),
{Phys. Rev.} {\bf D66},  014017 (2002) {Phys. Rev.} {\bf D66},  054005
(2002) 
I.W. Stewart, hep-ph/0208034 and references therein. 
M. Beneke et al., { Nucl. Phys.} {\bf B 643},  431 (2002) 
{ Phys. Lett.} {\bf B553},  267 (2003)




\end{thebibliography}
\end{document}